\documentclass[	
 						superscriptaddress,%
						twocolumn,%
						a4paper,
					]{revtex4-1}

\usepackage[frenchb,english]{babel}
\usepackage[T1]{fontenc}
\usepackage{upgreek}
\usepackage{textcomp} 				
\usepackage{graphicx}			
\usepackage{epstopdf}				
\usepackage{color}					
\usepackage{multirow}				
\usepackage[nice]{nicefrac} 		
\usepackage{amssymb,amsmath}		
\usepackage{array}					
\usepackage{ulem}
\usepackage{float}
\usepackage{tabularx}
\usepackage{natbib}
\usepackage{longtable}

\renewcommand{\emph}[1]{\textit{#1}}

\definecolor{vert}{rgb}{0.5,0.758,0.5}
\definecolor{bleufonce}{rgb}{0,0,0.516}
\definecolor{orange}{rgb}{1,0.516,0}

\begin{document}

\title{Archetypal soft-mode driven antipolar transition in francisite Cu$_3$Bi(SeO$_3$)$_2$O$_2$Cl}
\date{\today}

\author{Cosme Milesi-Brault}
\affiliation{Materials Research and Technology Department, Luxembourg Institute of Science and Technology, 41 rue du Brill, L-4422 Belvaux, Luxembourg}
\affiliation{University of Luxembourg, Physics and Materials Science Research Unit, 41 rue du Brill, L-4422 Belvaux, Luxembourg}
\author{Constance Toulouse}
\affiliation{University of Luxembourg, Physics and Materials Science Research Unit, 41 rue du Brill, L-4422 Belvaux, Luxembourg}
\author{Evan Constable}
\affiliation{Institute of Solid State Physics, Vienna University of Technology, 1040 Vienna, Austria}
\author{Hugo Aramberri}
\affiliation{Materials Research and Technology Department, Luxembourg Institute of Science and Technology, 41 rue du Brill, L-4422 Belvaux, Luxembourg}
\author{Virginie Simonet}
\author{Sophie de Brion}
\affiliation{Universit\'e Grenoble Alpes, CNRS, Institut N\'eel,38000 Grenoble, France} 
\author{Helmuth Berger}
\affiliation{Institute of Physics of Complex Matter, \'Ecole Polytechnique F\'ed\'erale de Lausanne, CH-1015 Lausanne, Switzerland}
\author{Luigi Paolasini}
\author{Alexei Bosak}
\affiliation{European Synchrotron Radiation Facility, BP 220, F-38043 Grenoble Cedex, France}
\author{Jorge \'I\~niguez}
\affiliation{Materials Research and Technology Department, Luxembourg Institute of Science and Technology, 41 rue du Brill, L-4422 Belvaux, Luxembourg}
\affiliation{University of Luxembourg, Physics and Materials Science Research Unit, 41 rue du Brill, L-4422 Belvaux, Luxembourg}
\author{Mael Guennou}
\affiliation{University of Luxembourg, Physics and Materials Science Research Unit, 41 rue du Brill, L-4422 Belvaux, Luxembourg}
\email{mael.guennou@uni.lu}

\begin{abstract}
Model materials are precious test cases for elementary theories and provide building blocks for the understanding of more complex cases. Here, we describe the lattice dynamics of the structural phase transition in francisite Cu$_3$Bi(SeO$_3$)$_2$O$_2$Cl at 115~K and show that it provides a rare archetype of a transition driven by a soft antipolar phonon mode. In the high-symmetry phase at high-temperatures, the soft mode is found at $(0,0,0.5)$ at the Brillouin zone boundary and is measured by inelastic X-ray scattering and thermal diffuse scattering. In the low-symmetry phase, this soft-mode is folded back onto the center of the Brillouin zone as a result of the doubling of the unit cell, and appears as a fully symmetric mode that can be tracked by Raman spectroscopy. On both sides of the transition, the mode energy squared follows a linear behaviour over a large temperature range. First-principles calculations reveal that, surprisingly, the flat phonon band calculated for the high-symmetry phase seems incompatible with the displacive character found experimentally. We discuss this unusual behavior in the context of an ideal Kittel model of an antiferroelectric transition.      
\end{abstract}


\maketitle

Antiferroelectric (AFE) materials undergo a structural phase transition from a high-symmetry phase to a low-symmetry variant where electric dipoles emerge or order, forming sublattices of opposite polarization that cancel out macroscopically~\cite{Rabe2013,Toledano2016}. Phenomenologically, AFE transitions are recognized as transitions between two non-polar phases that display an anomaly of the dielectric constant and where, in the low-symmetry phase, the application of an electric field leads to a phase transition towards a polar phase, which gives rise to the characteristic double hysteresis loop of the polarization as a function of field~\cite{Kanzig1957}. However, beyond phenomenology, there is no criterion to identify an antiferroelectric crystal structure, no consensus on a rigorous definition of antiferroelectric transitions~\cite{Rabe2013}, and the model of an elementary antiferroelectric material, as described already long ago by Kittel~\cite{Kittel1951}, has never found a practical realization in experiment. 

\begin{figure}[ht]
\includegraphics[width=0.4\textwidth]{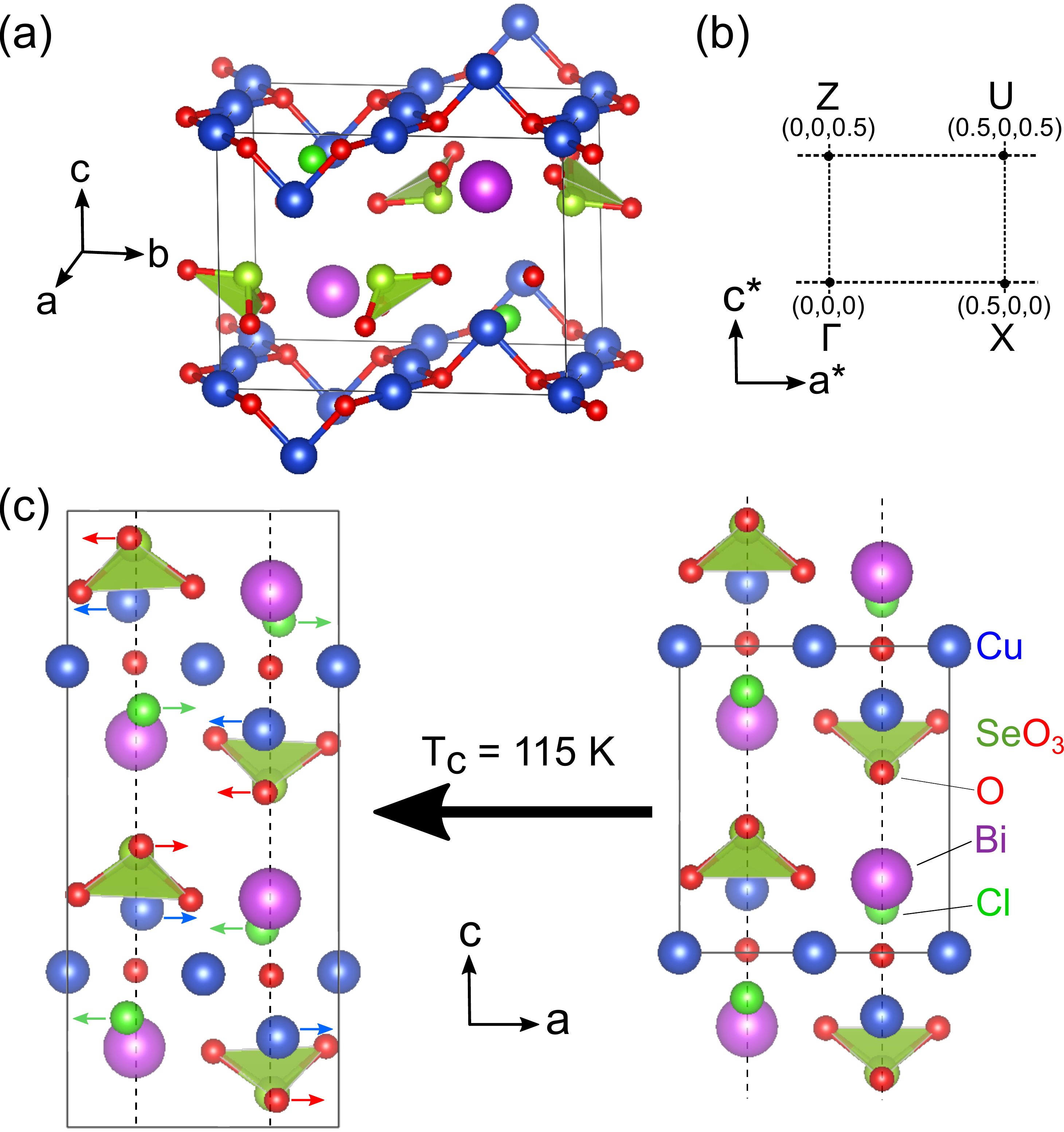}
\caption{(a) Crystal structure of francisite. (b) Sketch of the Brillouin zone with the labels of the zone-boundary points as used in Ref.~\cite{Aroyo2014}. (c) Crystal structure viewed along the $b$-axis, picturing the ionic displacements at the phase transition as measured by neutron diffraction.}
\label{fig:transition}
\end{figure}

In a direct analogy with ferroelectrics, it is natural to classify antiferroelectric transitions with respect to two idealized transition models: order-disorder and displacive~\cite{LinesAndGlass,Blinc1974,cummins1983}. Many AFEs can be clearly associated to the order-disorder type, like hydrogen-bonded crystals exemplified by squaric acid C$_4$O$_4$H$_2$~\cite{Horiuchi2018} or NH$_4$H$_2$PO$_4$~\cite{Fukami1987}, or systems where electric dipoles are localized on molecules like liquid crystals~\cite{Takezoe2010} or KCN~\cite{Stokes1984}. On the other hand, no example of a displacive antiferroelectric transition has been identified so far. Conceptually, it should be driven by an antipolar "soft" phonon mode, i.e. a structural instability at the boundary of the Brillouin zone involving only antiparallel ion displacements~\cite{Scott1974,Roos1976}, directly connected via the phonon branch to a polar mode at the zone center with in-phase motions of the same ions. As temperature goes down, the frequency of the antipolar phonon decreases and the ionic displacements eventually freeze at $T_c$, lowering the symmetry into the AFE structure, where the polar instability can be revealed by an electric field. In practice, such a transition hasn't been found and the concept of an "AFE soft mode" has been only used in the literature in complex situations where mode softening is not the primary driving force for the transition~\cite{Meister1969,Samara1970,Swainson2009,Hlinka2011a,Girard2018}. This in particular applies to the model AFE perovskite PbZrO$_3$~\cite{Shirane1951}, whose behaviour has been debated in recent years~\cite{Tagantsev2013,Hlinka2014,Iniguez2014,Burkovsky2017,Xu2019} and cannot be reduced to the softening of a single phonon mode. 

In this letter, we show that the orthorhombic francisite Cu$_3$Bi(SeO$_3$)$_2$O$_2$Cl (CBSCl) provides us with a remarkable case study that realizes the first clear example of a simple soft-mode driven antipolar transition that fulfills this condition. This compound, together with other related francisites, has been studied for some time for its pseudo-kagome lattice and its frustrated magnetism, and exhibits an antiferromagnetic order below $T_\mathrm{N}$ = 25~K~\cite{Pring1990,Millet2001,Miller2012,Pregelj2012,Rousochatzakis2015,Constable2017}. In addition, it also undergoes a structural phase transition at 115~K, characterized by antiparallel displacements of the Cl and Cu ions along the $a$ axis, leading to a doubling of the unit cell along the $c$ axis (Fig.~\ref{fig:transition}). Neutron refinements have established a symmetry descent $Pmmn\longrightarrow Pcmn$, whereby a mirror plane is replaced with a glide plane~\cite{Constable2017}, which corresponds to the most simple one-dimensional AFE transition~\cite{Toledano2016}. The dielectric anomaly~\cite{Constable2017} and the nearly degenerate energies of the non-polar and polar polymorphs~\cite{Prishchenko2017} very strongly suggest antiferroelectric properties, even though the phase transition under electric field has not yet been demonstrated. Here, our study is focused on its lattice dynamics at low frequencies, with the aim to elucidate its driving mechanism.

\begin{figure}
\includegraphics[width=0.5\textwidth]{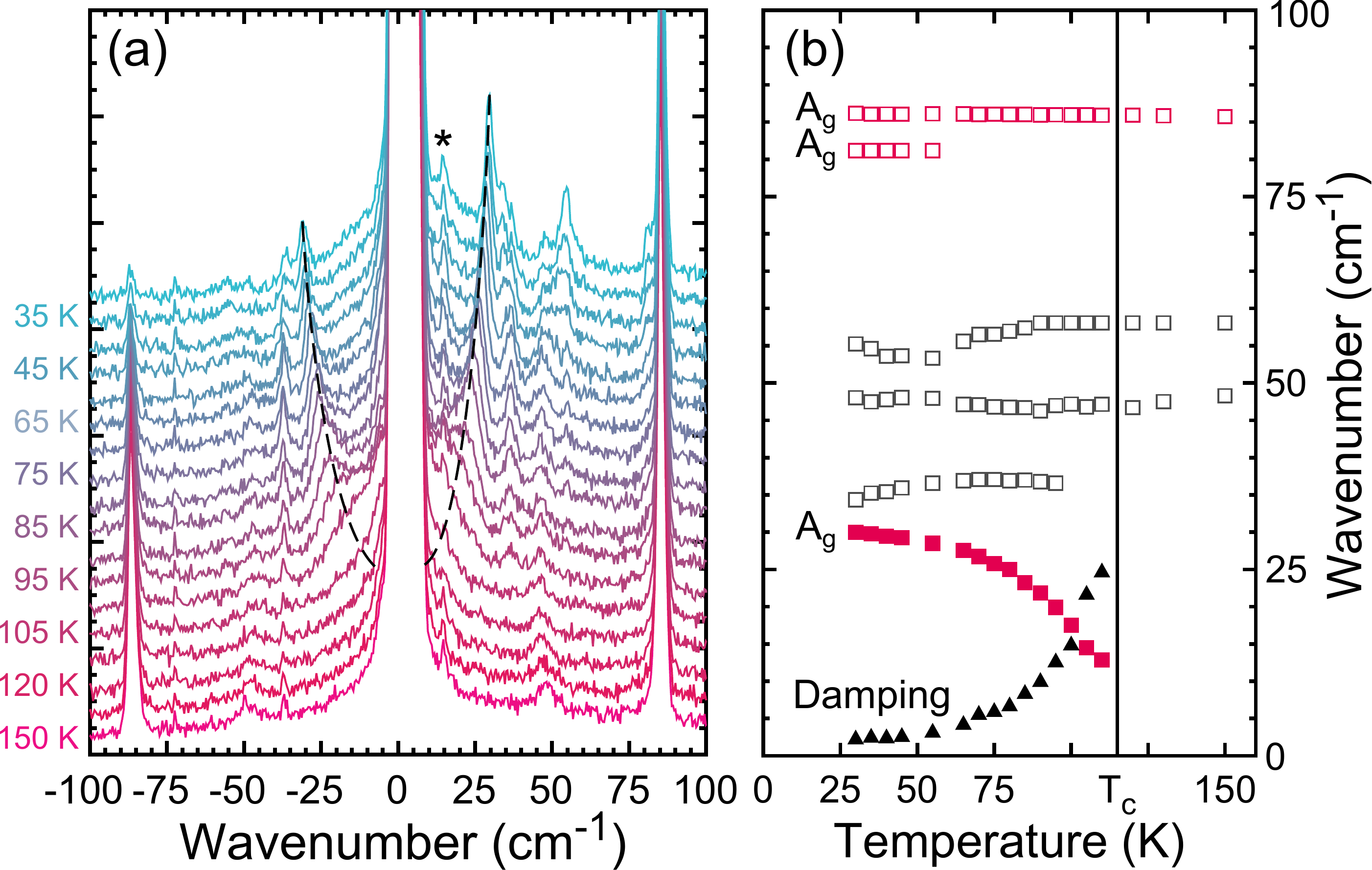}
\caption{(a) Low-frequency Stokes$-$Anti-Stokes Raman spectra as a function of temperature across the phase transition. The dashed line is a guide to the eye for the behavior of the soft mode. The peak marked with an asterisk * is an artefact. (b) Mode frequencies as a function of temperature. The soft-mode data are pictured with full symbols, including its damping shown as black triangles.}
\label{fig:raman}
\end{figure}

We first show the phase transition as seen in Raman spectroscopy. The crystal investigated here is similar to those used in~\cite{Constable2017}. The layered crystal structure depicted in Fig.~\ref{fig:transition} favors growth along the $[010]$ direction and the formation of thin plate-like crystals in the $ab$ plane, with the $c$ axis normal to the sample faces, which in Raman spectroscopy gives access to the $A_g$ and $B_{1g}$ phonon modes. The sample was placed in an Oxford Microstat-Hire open-cycle He cryostat and the Raman spectra were measured with a 633~nm He-Ne laser in an inVia Renishaw micro-Raman spectrometer using Bragg filters with a cut-off at 10~cm$^{-1}$. Fig.~\ref{fig:raman} (a) shows the Raman spectra at different temperatures, starting slightly above $T_\mathrm N$ so as to remain in the paramagnetic phase at all temperatures and avoid the anomalies related to the magnetic transition~\cite{Gnezdilov2017}. At 30~K, the very low frequency range is dominated by an intense mode at 36~cm$^{-1}$ that continuously goes down in frequency as temperature is increased, vanishes at the transition, and is not visible above $T_c$. We identify it as the soft mode driving the transition. In addition, the transition is also revealed by the emergence, below $T_c = 115$~K, of some hard phonon modes, consistent with previous reports~\cite{Gnezdilov2017}. Raman modes were fitted with damped harmonic oscillators and their frequencies are reported in Fig.~\ref{fig:raman} (b). The lowest frequency mode exhibits a very rapid decrease in frequency, and remains underdamped until ~15~K below the transition. The other (hard) phonon modes, on the other hand, are nearly insensitive to the phase transition. This observation is consistent with two scenarios. In the first scenario, the transition is ferroelectric and the soft mode is a polar mode that is visible in Raman spectroscopy only in the polar phase. In the second scenario, the transition is between two non-polar phases, with a doubling of the unit cell, and the soft mode is found at the zone boundary in the high-symmetry phase, so that it can no longer be followed by optical spectroscopy techniques. Crystallographic data~\cite{Constable2017} and the absence of a soft mode in infrared spectroscopy~\cite{Miller2012} support the latter scenario, but the former has also been discussed in past spectroscopic studies~\cite{Gnezdilov2016,Gnezdilov2017}. 

\begin{figure*}
\includegraphics[width=0.95\textwidth]{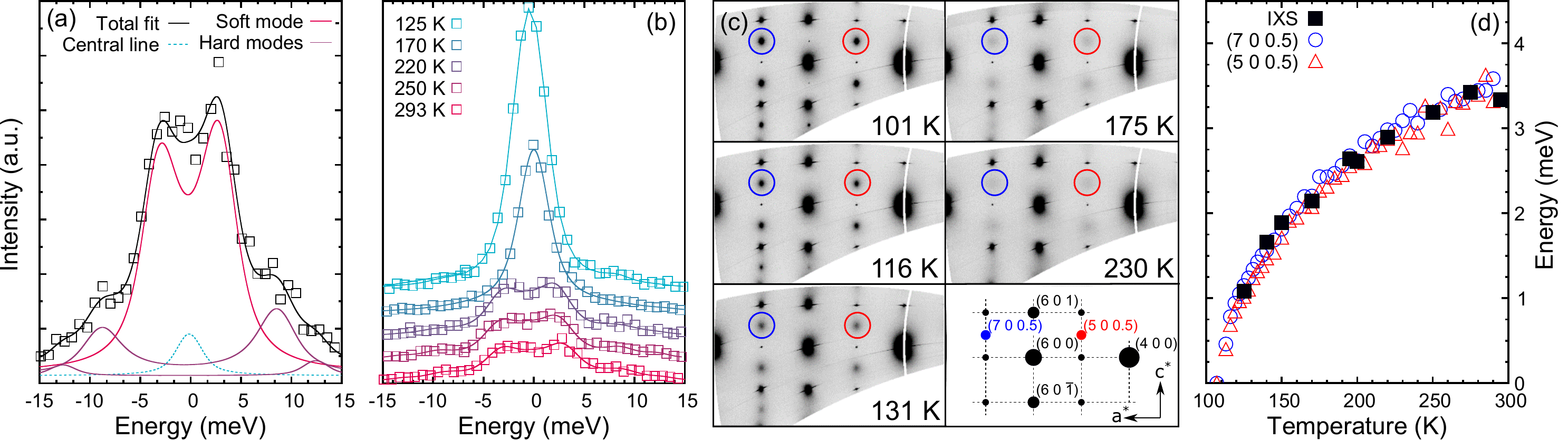}
\caption{(a) Representative IXS spectrum at room temperature at $(5,0,0.5)$, with the fit with damped harmonic oscillators. (b) Temperature evolution of the IXS spectrum from room temperature to $T_c$. (c) Reconstructed ($h$0$l$) layer of francisite reciprocal space at selected temperatures. (d) Evolution of the fitted soft-mode with temperature using both data sets.} 
\label{fig:INX1}
\end{figure*}

In order to confirm whether this comes with a mode softening at the zone boundary above $T_c$ as anticipated from the structural studies, we performed inelastic X-ray scattering (IXS) to follow the mode at the $Z$ point of the Brillouin zone (See Fig.~\ref{fig:transition}). The measurements were performed on the ID28 beamline at the European Synchrotron Radiation Facility (ESRF), as described elsewhere~\cite{Krisch2007}. A typical energy scan at room temperature at $(5,0,0.5)$ is shown in Fig.~\ref{fig:INX1} (a). The spectrum is dominated by a low-energy phonon around 3~meV which we identify as the soft mode. Some weaker modes are clearly present at higher energies. We fitted the spectra with the soft mode, two hard modes and a weak central line. It is clear from the Raman results that more than two hard modes may be present in this region, but the experimental resolution does not allow us to resolve them, and we do not make any attempts at assigning them precisely. The dispersion was also measured along the $Z$--$\Gamma$ and $Z$--$U$ directions (Supplementary information). They both indicate that the mode hardens when moving away from the $Z$ point, and belong to an optic branch. As temperature is decreased from room temperature towards the phase transition, the dominant mode shows again a typical soft-mode behaviour~\cite{cummins1983}, with a marked softening and an increase in intensity. The hard modes have, as expected, negligible contributions to the intensity and moderate temperature variations, while the soft mode frequency goes down. Fits with damped harmonic oscillators could be performed down to 125~K below which it becomes resolution limited. The damping did not show any significant change in this range within the resolution of the experiment. 

In order to continue following the soft-mode frequency to the lowest possible energies, we investigated the temperature dependence of thermal diffuse scattering (TDS) on the same beamline~\cite{Girard2019}. Under the assumption that TDS comes predominantly from a single phonon, which is amply justified by the IXS spectra shown above, the intensity is related to that phonon frequency by the expression~\cite{Girard2018}
\[
I_\mathrm{TDS} \approx A\frac{1}{\omega}\coth\left(\frac{\hbar\omega}{2k_\mathrm BT}\right)
\]
where $\omega$ is the frequency of the phonon, $\hbar$ the reduced Planck constant, $k_\mathrm{B}$ the Boltzmann constant and $T$ the temperature. $A$ is here a constant prefactor that was dealt with by rescaling the whole evolution in order to match the absolute values found by IXS. Fig.~\ref{fig:INX1} (c) shows reciprocal space maps at selected temperatures, showing the increase in intensity of the scattering, and Fig.~\ref{fig:INX1} (d) shows the evolution of the soft-mode energy derived from the scattering intensities at the $(5,0,0.5)$ and $(7,0,0.5)$ reciprocal lattice positions. Both evolutions are similar and after rescaling nicely interpolate the values found from the IXS fits over the full temperature range, which strengthen the relevance of the hypothesis used for the analysis of TDS.

\begin{figure}[t]
\includegraphics[width=0.4\textwidth]{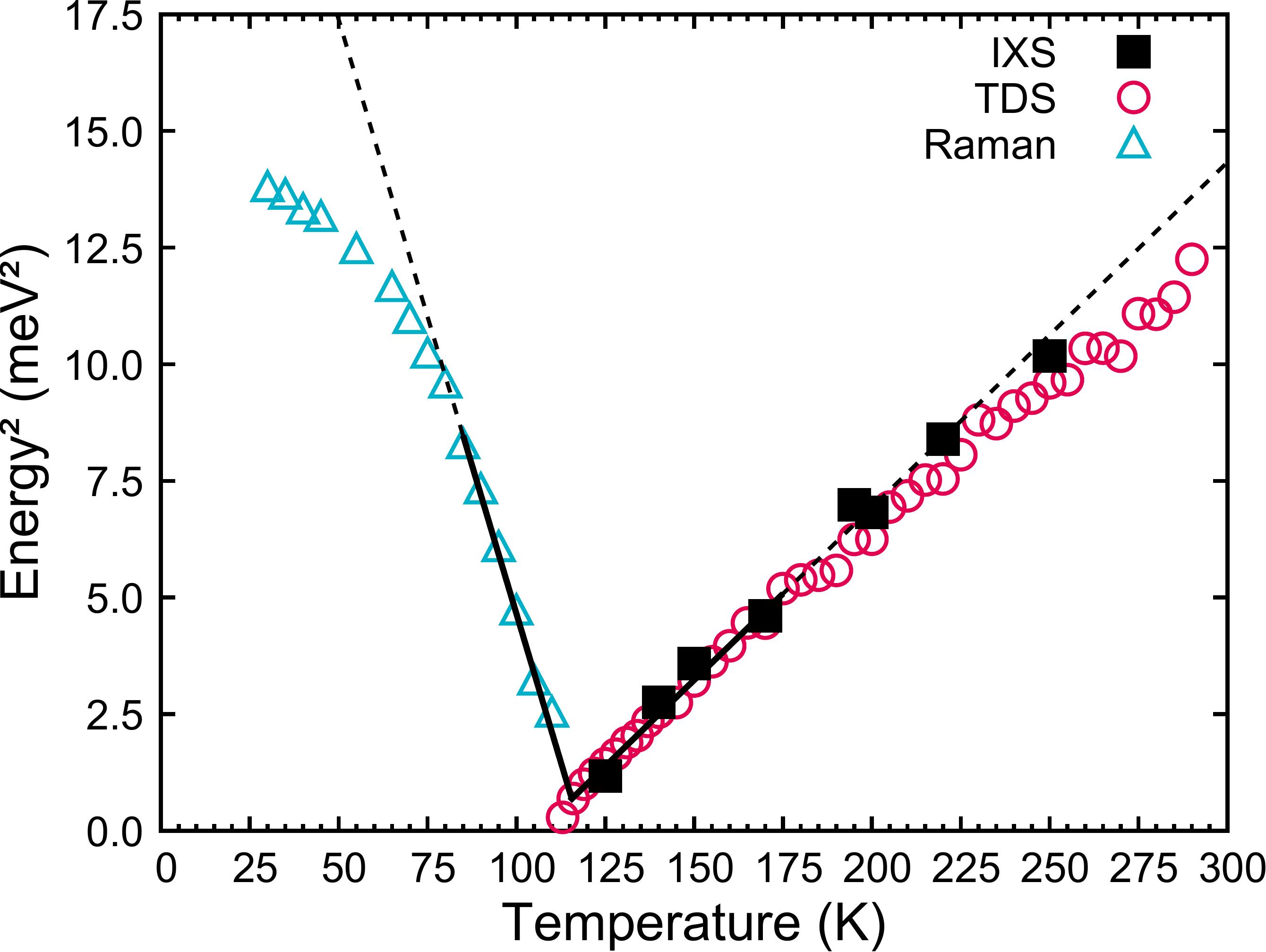}
\caption{Energy squared of the soft mode in both the low and high temperature phases. The solid line shows the region used for the linear fitting.} 
\label{fig:softmode}
\end{figure}

All experimental data from Raman spectroscopy, IXS and thermal diffuse scattering are summarized in Fig.~\ref{fig:softmode}. Following classical theory of soft mode transitions~\cite{Dove1997,Blinc1974}, we expect the squared energy of the soft mode to follow a linear behaviour in the vicinity of the transition. Fig.~\ref{fig:softmode} shows that this is indeed realized within a large temperature range. The expectation from the most simple Landau model, whereby the Landau potential is expanded up to the 4\textsuperscript{th} order in the order parameter as $G(Q) = a_0(T-T_c)Q^2/2 + bQ^4/4$, is that the ratio of the slopes $\partial E^2/\partial T$ above and below $T_c$ equals 2. Here, the linear regression showed in figure~\ref{fig:softmode} gives a value of 3.4, which indicates that a Landau model should at least include the 6\textsuperscript{th} order term $cQ^6/6$. This is intermediate between the values expected for a purely second-order transition (2) and a purely tricritical transition (4, for $b=0$ and $c\neq 0$). In that respect, this case is similar to the archetypal antiferrodistorsive transition in SrTiO$_3$~\cite{Salje1998,Carpenter2007}, which is also close to tricritical at ambient pressure. Finally, we observe in Fig.~\ref{fig:softmode} deviations from linearity far above and far below the transition temperature. We attribute the high-temperature deviations predominantly to mode coupling mechanisms (described below). At low temperatures, we note that magnetic correlations develop at about 75~K, characterized by nearest neighbor interactions of the order to 6.6~meV~\cite{Constable2017}, meaning that spin-lattice coupling may play a role in the deviations.

\begin{figure}
\includegraphics[width=0.48\textwidth]{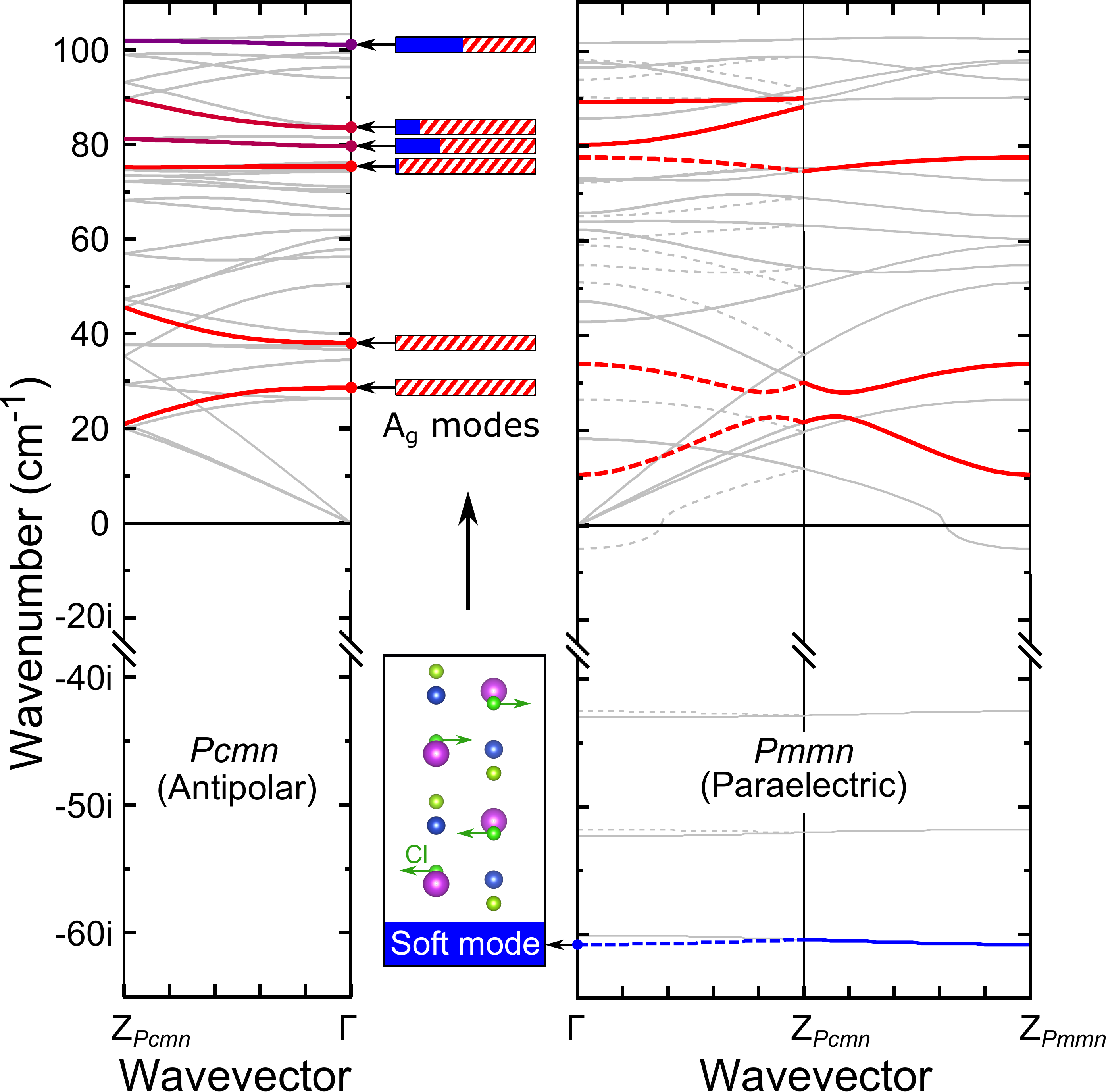}
\caption{Calculated phonon dispersion relations along the $\Gamma$--$Z$ line in the low-symmetry antipolar $Pcmn$ phase (left) and the high-symmetry paraelectric $Pmmn$ phase (right). For the $Pmmn$ phase, the branches have also been folded for an easier comparison with the $Pcmn$ phase (dashed lines). The red thicker lines are the branches originating from the fully symmetric Raman-active $A_g$ modes that are relevant to the present discussion. The middle panel shows how the unstable soft mode calculated in the high-symmetry phase projects onto the $A_g$ modes in the antipolar phase, where the (solid) blue and (hashed) red colors show the contributions of the soft mode and the other modes, respectively.} 
\label{fig:dispersion}
\end{figure}

Lattice dynamical calculations were performed by density functional theory (DFT) as implemented in the VASP code (further details are given in the Supplementary Information). The phonon modes and their dispersion branches were calculated for the high-symmetry ($Pmmn$) and the low-symmetry ($Pcmn$) phases (Fig.~\ref{fig:dispersion}). While the $Pcmn$ phase is dynamically stable, the high-symmetry phase possesses several unstable modes showing up as imaginary frequencies, consistent with previous findings~\cite{Gnezdilov2017}. The most unstable modes are the ones driving the transition and are found along the $\Gamma$--$Z$ direction, whereby their branch is nearly flat, with the phonon mode at $Z$ only slightly lower than the zone-center phonon. This is consistent with past studies~\cite{Prishchenko2017}, and strongly supports the potential antiferroelectric nature of the transition~\cite{Rabe2013}.

The DFT results also give some more insight into the mode coupling mechanisms. As it is apparent in Fig.~\ref{fig:dispersion} from the comparisons of the low-lying phonon modes, there is no obvious mapping between the phonons and branches of the two phases. A careful check, by projecting the soft mode eigenvector onto the phonons of the low-symmetry phase, reveals that the original soft mode with Cl displacements is distributed mainly on three hard phonon modes above 75~cm$^{-1}$. On the other hand, the lowest $A_g$ mode in the AFE phase, which we assign to the soft mode observed in Raman spectroscopy, includes displacements of Bi, Cu, Se and Cl, none of which clearly dominates. The simple image of the transition being only driven by the antipolar displacements of the Cl ions is therefore an approximation only. In fact, mode coupling is inevitable in this case: one of the zone-boundary phonons connected to the acoustic branch must have the same symmetry ($Z_3^-$) as the antipolar soft mode, and will mix with it as soon as it shifts upwards. This is a major difference with the archetypal antiferrodistorsive transitions in SrTiO$_3$ or LaAlO$_3$, where the fully symmetric soft mode is not forced to couple with other low-lying modes and can preserve its simple ionic displacement pattern~\cite{Fleury1968,Scott1969}. 

On the other hand, this calculation is surprisingly inconsistent with the experimental observations on the nature of the transition: a dispersionless phonon branch as seen here in Fig.~\ref{fig:dispersion} usually implies an order-disorder behavior, while a displacive transition -- and the validity of the simple Landau approach used above -- requires that the instability be restricted to a small region in reciprocal space. In other words, this is seemingly a very unusual case where phonon bands calculated by DFT at 0~K for the high-symmetry phase do not reflect the nature of the phonon bands at temperatures above the phase transition. The full anharmonic model needed to correctly describe the temperature effects is beyond the scope of this paper; but it will probably modify the details of the ionic motions involved in the soft mode. More importantly, this exception points to a difficulty pertaining to displacive Kittel-like antiferroelectric transitions in general. The near-degeneracy of the polar and antipolar phases is required for switching under field to occur. This requires that the polar and antipolar soft modes should be also close in energy, which is difficult to combine with strong curvatures of the phonon branch at either ends, unless a large anharmonicity (and strong temperature-dependence of the bands) comes into play. This makes this simple transition in fact non trivial and highlights its interest as a model system. This also makes it even more crucial to verify the switching under electric field, both experimentally and theoretically, and the behaviour of low-frequency polar phonon modes, in order to ultimately validate the antiferroelectric -- and not only antipolar -- properties of francisite.

In summary, we have shown that an archetypal soft-mode driven antipolar transition is realized in francisite at 115~K, which makes it the closest example known to date of an elementary, Kittel-like, transition. This finding opens a playground to verify elementary theories on displacive antiferroelectrics for which no antiferromagnetic equivalent exists. Besides, given the rich magnetism found in francisite at lower temperatures, opportunities for studying the coupling between elementary AFE behaviour and magnetism in general are also opened. Magnetoelectric coupling in francisite has recently been established in the static regime~\cite{Constable2017}. It is therefore conceivable that the dynamic components of the magnetoelectric susceptibility could support novel excitations analogous to the electromagnons found in conventional multiferroics~\cite{Pimenov2006}, the so-called "AFE-electromagnon". The confirmation of francisite as a proper AFE material will be the first step in the hunt for these new magnetoelectric excitations.

\begin{acknowledgments}
C.~M.-B., C.~T., H.~A., J.~I. and M.~G. acknowledge financial support by the Luxembourg National Research Fund under projects C16/MS/11348912/BIAFET/Guennou and INTER/ANR/16/11562984/EXPAND/Kreisel.
\end{acknowledgments}


\clearpage

\end{document}